\documentstyle[aps,psfig,multicol]{revtex}
\begin{document}
\setlength{\topmargin}{-1.3cm}
\draft
\title{Evolving networks with disadvantaged long-range connections}
\author{R. Xulvi-Brunet$^1$ and I.M. Sokolov$^{1,2}$}
\address{$^1$Institut f\"{u}r Physik, Humboldt Universit\"{a}t zu Berlin, 
         Invalidenstra\ss e 110, D-10115 Berlin, Germany}
\address{$^2$Theoretische Polymerphysik, Universit\"at Freiburg, Hermann Herder 
         Str. 3, D-79104 Freiburg, Germany}
\date{\today}
\maketitle

\bigskip

\begin{abstract}
{\bf Abstract.} We consider a growing network, whose growth algorithm is based 
on the preferential attachment typical for scale-free constructions, but where 
the long-range bonds are disadvantaged. Thus, the probability to get connected 
to a site at distance $d$ is proportional to $d^{-\alpha }$, where $\alpha $
is a tunable parameter of the model. We show that the properties of the
networks grown with $\alpha <1$ are close to those of the genuine scale-free
construction, while for $\alpha >1$ the structure of the network is vastly 
different. Thus, in this regime, the node degree distribution is no more a 
power law, and it is well-represented by a stretched exponential. On the 
other hand, the small-world property of the growing networks is preserved at 
all values of $\alpha $.
\end{abstract}


\begin{multicols}{2}

Complex weblike structures (the small-world or scale-free networks) have 
recently become an object of extensive investigation, and in the last years a 
great success in understanding the properties of these structures was achieved 
(see Ref.\cite{review} as a review). Apart from appealing mathematics, 
this recent interest is due to the fact that many natural and technological 
systems, like polymer networks \cite{Sokolov}, the science collaboration network 
\cite{New1,New2,New3}, or  networks of chemical reactions in a living cell 
\cite{Oltvai,Fell,Mason} seem to be organized according to some internal 
principles. Thus, the Internet \cite{Faloutsos}, the network of human sexual 
contacts \cite{Amaral} or the WWW \cite{Albert} possess a similar structure, e.g. 
are they all based on the preferential attachment of the newly introduced nodes 
to the highly connected old ones. All these networks show the small-world 
property: the typical distance (in terms of the number of intermediate 
connections) between two nodes grows logarithmically with the web's size. 

One of the prominent examples of a mathematical model of such a growing network 
is the scale-free (SF) construction of Barab\'{a}si and Albert 
\cite{review,scalefr}; and one of its most interesting properties is the 
very specific form of the probability distribution of the degree of 
nodes (i.e. of the numbers of bonds connecting any given node $i$ with other 
ones in the network): $P(k)\propto k^{-3}$ \cite{review,scalefr,Do2,Kull,Kra}. 
Many models have been presented, based on the same two most important 
ingredients: growth and preferential attachment. Examples are models with an 
accelerated growth of the network \cite{Do4,Bar3}, models with a nonlinear 
preferential attachment \cite{Kra}, with nodes provided by a initial 
attractiveness \cite{Do2,Je1}, with growth constraints as aging and cost 
\cite{Amar,Do1}, models that have a competitive aspect of the nodes 
\cite{Bianconi}, or models of networks that incorporate local events as the 
addition, rewiring or removal of nodes or edges \cite{Alb}.

The SF-construction may be a reasonable approximation for such
world-spanning networks like one of the Internet's information transmission 
channels or one of the formal links of WWW. On the other hand, in many
situations (like in a network of human sexual contacts) a connection
means a physical contact, i.e. means that the contacting individuals,
representing the nodes of the network, have to occur at the same site and at 
the same time, thus introducing a clear geographical aspect. In what follows 
we present a simple model taking into account this metrical (''geographical'')
aspect, where the probability to connect two nodes depends both on the number 
of connections that the nodes already have (as in the genuine SF-construction), 
and on the distance between them. That is, we treat an emerging network in a 
metric space. In this emerging network the probability that a newly introduced 
node $n$ is connected to a previously existing node $i$ is proportional to the 
number $k_{i}$ of the already existing connections of node $i$ (preferential 
attachment prescription), but on the other hand the too long bonds are 
disadvantaged, because this probability depends on the Euclidean distance 
$d_{in}$ between the nodes $n$ and $i$ as $d_{in}^{-\alpha }$, (clearly, a 
''scale-free'' function), with $\alpha >0$.

Based on extensive numerical simulations of a one-dimensional situation, we
show that even if the length penalties are mild, the model exhibits
properties which differ strongly from those of the usual scale-free networks.
Thus, the corresponding degree distribution function $P(k)$ depends strongly
on $\alpha $. We show, in particular, that for $\alpha <1$ the behavior of 
$P(k)$ is similar to the behavior of the SF model without penalties, so that 
asymptotically $P(k)\propto k^{-3}$, (which distribution possesses a mean, but 
no dispersion, and corresponds to strong, universal fluctuations). On the other
hand, for $\alpha >1$ the behavior of $P(k)$ is well-described by a
stretched-exponential $P(k)\propto \exp (-bk^{\gamma })$, with the power $%
\gamma $ depending on $\alpha $, so that the fluctuations in $k$ are rather
weak. We discuss the reasons for such a dramatic change, being rooted in the
probability of connection between the nodes as function of the distance, and
the overall structure of the emerging network, preserving its small-world
nature even at large (probably at all) $\alpha $-values.

\end{multicols}

\begin{figure}[t]
  \centerline{\hspace{-1cm}\psfig{figure=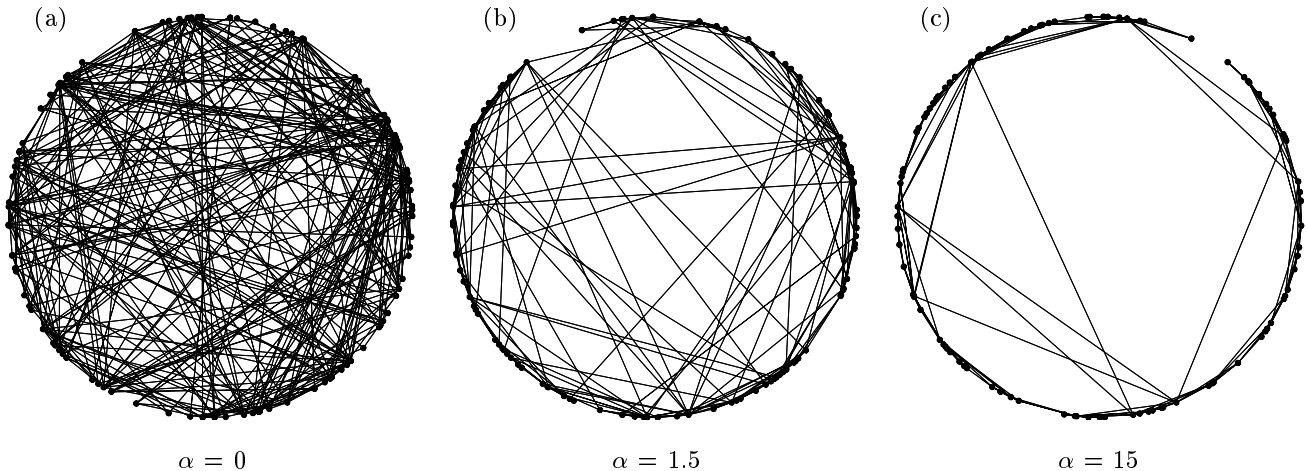,width=9in}}
  \caption{Networks generated using the simulation prescription, Eq.(2), with 
           different values of $\alpha$: (a) $\alpha=0$, (b) $\alpha=1.5$ 
           and (c) $\alpha=15$. All three examples have $300$ edges, $L=10^6$, 
           $N=102$, and $m=3$. Note the change in the appearence of the 
           networks. The network (a) is a genuine SF construction while (c) 
           strongly resembles the Watts and Strogatz's small-world network.}
  \label{fig:5}
\end{figure} 

\bigskip

\begin{multicols}{2}

We start from a one-dimensional lattice of $L$ sites, spaced by a unit
distance and apply cyclic boundary conditions. On this structure we will let 
our network grow, so that each lattice site will be a possible location of a 
network's node. We denote by $n_{i}$ the position in the lattice of a node $i$. 
The distance $d_{ij}$ between any two nodes $i$ and $j$ is defined as:

\begin{equation}
 d_{ij}=min\{|n_{i}-n_{j}|,(L-|n_{i}-n_{j}|)\}.
\end{equation}

Let us now construct the network. First, we choose randomly an even number $%
m_{0}$ of sites from the lattice and we bind them in pairs with one bond each.
This will be our initial condition. That is, at $t=0$, our network will
consist from $m_{0}$ nodes connected in pairs. As in the SF model we will
add at every time step a new node to our network (linear growth). We proceed
according to the following rule: at every time step we choose at random a
free site of our lattice, and pose the new node there. This new node is then
connected through $m$ edges ($m\le m_{0}$) with $m$ different nodes already
present in the network. After $t$ time steps the algorithm results
in a network with $t+m_{0}$ nodes and $mt+m_{0}/2$ edges. In contrast with
the SF model, the probability $\Pi $ for the new node $n$ to be connected to
an old one $i$ will depends not only on the number of edges $k_{i}$ which $i$
already possesses, but also on the distance $d_{in}$ between them:

\begin{equation}
\Pi (k_{i},d_{in},\alpha )=\frac{k_{i}\cdot d_{in}^{-\alpha }}{%
\sum_{j}k_{j}\cdot d_{jn}^{-\alpha }}.  \label{Apriori}
\end{equation}

Here the sum in the denominator goes over all nodes in the system except the
newly introduced one and $\alpha $ is a real non-negative parameter
describing the distance penalties. For large $\alpha $, the probability of
connection between two distant nodes is very small. On the other hand,
for a very small $\alpha $ the probability is almost independent from the
distance. In the case $\alpha =0$ our model reduces to the genuine
scale-free one. Note that our model is to some extent also scale-free: the
connection probabilities depend only on the {\it relative} distances.

Our initial condition is slightly different from one of Barab\'{a}si and
Albert, where the initial $m_{0}$ nodes are not connected: in our case all
nodes introduced at $t=0$ have exactly one edge, which allows to use Eq.(\ref
{Apriori}) from the very beginning. This simplifies the algorithm, since we
do not have to distinguish between the initial and the further steps. The
only difference with the genuine SF construction is that at time $t$ one has 
$mt+m_{0}/2$ (instead of $mt$) edges present; hence, the asymptotic behavior of
both models for $t\rightarrow \infty $ is the same. 

Three examples of the evolving networks of such a kind are given en Fig.1. Here
is $m=3$, $L=10^6$, $N=102$ and $m_0=6$, (so that all three networks have 
exactly $300$ edges). Three different values of $\alpha$ were used: 
$\alpha=0.0$ (scale-free model), $\alpha=1.5$ and $\alpha=15.0$. Note that 
increasing $\alpha$ leads to marked changes in the topology of the network. 
Fig. 1(a) corresponds to a genuine scale-free construction and exhibits a 
lot of long bonds connecting distance sites. On the other hand, only few such 
bonds are present in Fig. 1(c).

In our further simulations we use a lattice of $L=2\cdot 10^{7}$ 
sites; the maximal number of the introduced nodes is $N=2\cdot 10^{5}$. All 
simulation results are based on the average of 10 realizations of this 
structure. The error bars on Figs. 3-5 correspond just to this ensemble average. 
The simulations are done for several values of $\alpha $ and for two values of 
$m$, the number of the outgoing bonds: $m=1$ and $m=3$; $m_0=2m$. 
  
\begin{figure}[htbp]
  \centerline{\hspace{0.4cm}\psfig{figure=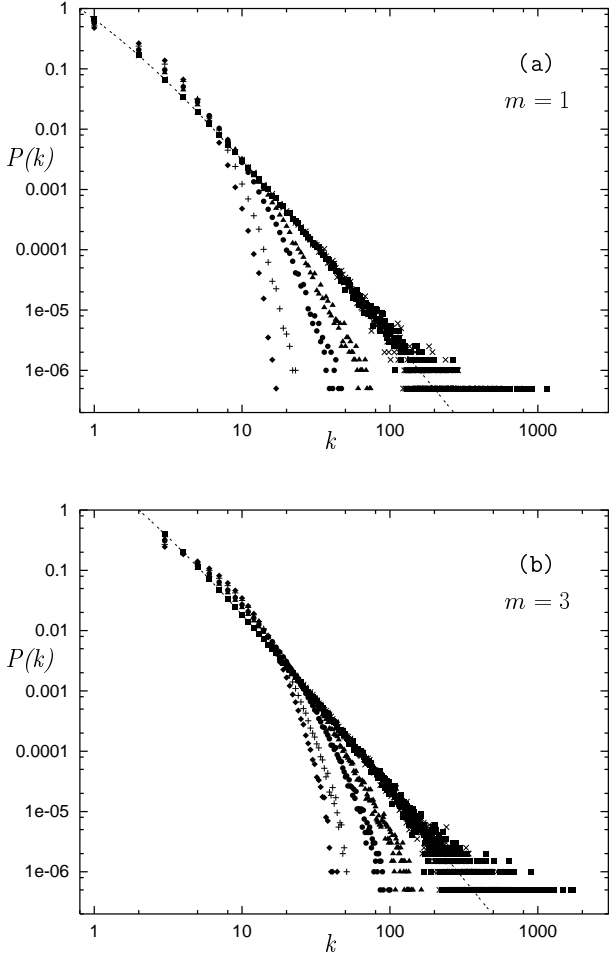,width=3.9in}}
  \caption{The degree distribution $P(k)$ for different values of $\alpha$ and
           for $m=1$ (a) and $m=3$ (b). The values of $\alpha$ are 
           $\alpha=0$ (squares), $\alpha=0.8$ (crosses), $\alpha=1.5$ 
           (triangles), $\alpha=2$ (filled circles), $\alpha=5$ (plusses) 
           and $\alpha=45$ (diamonds). The dashed lines correspond to the 
           theoretical curve for the scale-free model, (Ref. [1,12])}. 
           
  \label{fig:5}
\end{figure}
\bigskip

One of the prominent features of the scale free-model is that the
distribution of the degrees of the nodes decays as a power law, i.e. like $%
P(k)\sim k^{-\gamma }$, with $\gamma =3$. This corresponds to the fact that
the mean number of connections per site exists, but its dispersion diverges.
Let us now discuss, how this distribution changes if the long-range
connections are penalized. In Fig. 2 we plot the probability distribution of 
$k$ for different values of $\alpha $ on double logarithmic scales. One
readily infers that for all $0<\alpha <1$ no important differences with the
scale free model ($\alpha =0$) can be detected: in any case the
asymptotic behavior of $P(k)$ is well-described by $P(k)\sim k^{-3}$. The
distributions seem to be almost identical; however, small but statistically
significant deviations can be detected for small $k$-values. At $\alpha
\simeq 1$ the degree distribution shows a pronounced change in its behavior
and ceases being a power law; now the behavior of the model with distance
penalties is quite different.

Let us concentrate on the case $\alpha >1$ and try to describe the shape
of the degree distribution under such conditions. The analysis of the
simulations suggests that the corresponding mathematical expression could be
a stretched-exponential function of the form:

\begin{equation}
P(k)=a\exp (-b\,k^{\gamma }) ,  \label{Fit}
\end{equation}
where the parameters $a$, $b$ and $\gamma $ depend on $\alpha $ and $m$. To
obtain the values of these parameter and to analyze the goodness of this
fitting function we have fitted the data to Eq.(\ref{Fit}) using the nonlinear 
least-squares Levenberg-Marquardt algorithm \cite{Marq}, taking into 
consideration the error bars as coming out of $10$ realizations of each 
situation. The data is replotted together with the outcomes of the fits in 
Fig. 4 on the scales in which the fitting function, Eq.(\ref{Fit}), is 
represented by a straight line. One namely takes $k^{\gamma }$ as the abscissa 
and $\ln P(k)$ as the ordinata of the graph. Fig. 4 shows that such a fit 
(straight line) is surprisingly good!

The values of the exponent $\gamma $ are 
shown as a function of $\alpha $ ($\alpha >1$) in Fig. 3, for the two different 
situations corresponding to $m=1$ and $m=3$. We see that $\gamma $ 
monotonously grows with $\alpha $, and that the dependences $m=1$ and $m=3$ 
differ, i.e. that the $\gamma (\alpha )$ dependence is nonuniversal.

\begin{figure}[htbp]
  \centerline{\hspace{-0.4cm}\psfig{figure=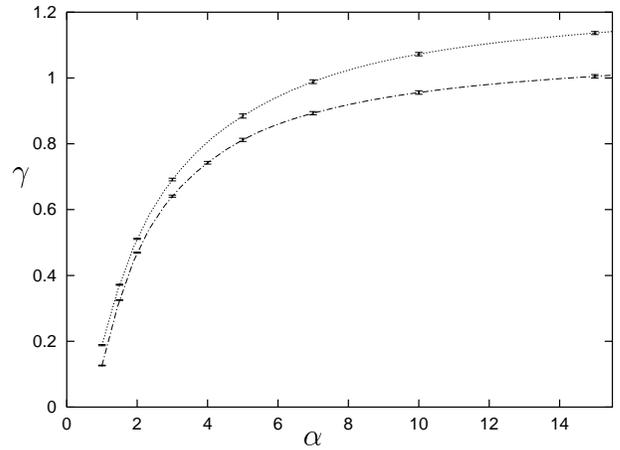,width=4.75in}}
  \caption{The parameter $\gamma$ as a function of $\alpha$. The upper 
           dependence corresponds to $m=1$, and the lower one $m=3$. The 
           lines are drawn as a guide for eyes.}
  \label{fig:5}
\end{figure}
\bigskip

We note that in related models of growing networks another form of 
degree distribution appears: an exponentially demped power-law \cite{Newman},

\end{multicols}

\begin{figure}[t]
  \centerline{\hspace{-0.2cm}\psfig{figure=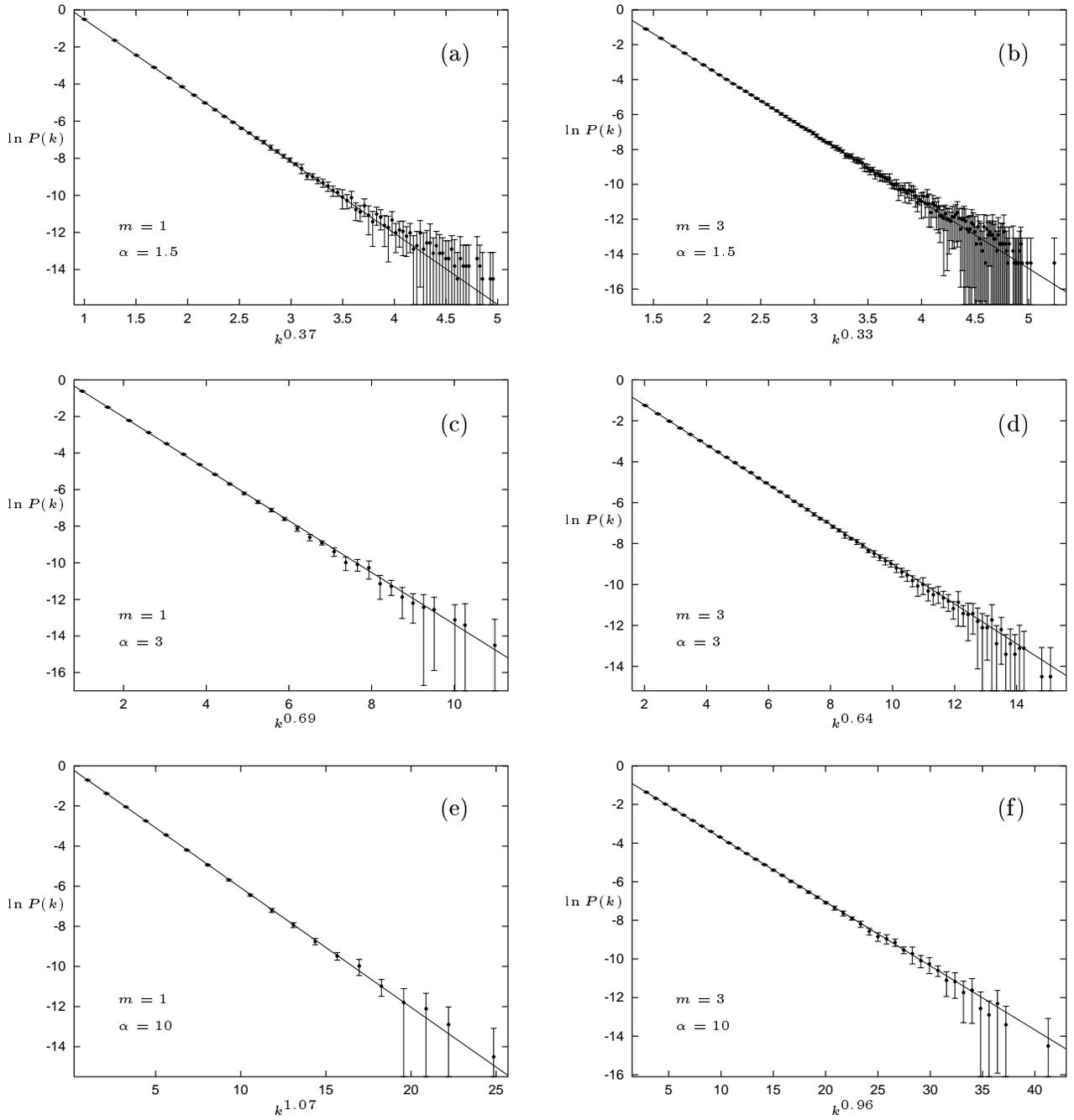,width=10in}}
  \caption{Shown is $\ln{P(k)}$ as a function of $k^{\gamma}$, where $\gamma$ 
           is the output of the fit, Eq.(3). (See text for details). The 
           parameters are: (a): $m=1$, $\alpha=1.5$, $\gamma=0.37$. (b): 
           $m=3$, $\alpha=1.5$, $\gamma=0.33$. (c): $m=1$, $\alpha=3$, 
           $\gamma=0.69$. (d): $m=3$, $\alpha=3$, $\gamma=0.64$. (e): $m=1$, 
           $\alpha=10$, $\gamma=1.07$. (f): $m=3$, $\alpha=10$, $\gamma=0.96$.}
  \label{fig:5}
\end{figure}
\bigskip

\begin{multicols}{2}

\begin{equation}
P(k)=ak^{\gamma }\exp (-b\,k).  \label{Cut}
\end{equation}
We tested also this fit function and found out that it gives a good fit for
larger $\alpha $-values, but is definitely inferior to our fit, Eq.(\ref{Fit}), 
for $1<\alpha <3$.

A growing network with disadvantaged long bonds is a very interesting
hierarchical construction. Thus, for large $\alpha $, the strong correlation
between the age of the connection and its length exists. The old
connections, made when the nodes were sparce, are typically long, while the
younger connections get shorter and shorter, since more sites in the
immediate vicinity of a newly introduced site can be found. The simulations
show that for $\alpha $ large, the nodes are almost surely connected to
their nearest neighbors. On the other hand, the old, long-range connections
are of great importance for the overall topology of the lattice, since
they guarantee that for any $\alpha $ the network is a small-world one. 

\begin{figure}[htbp]
  \centerline{\hspace{-0.3cm}\psfig{figure=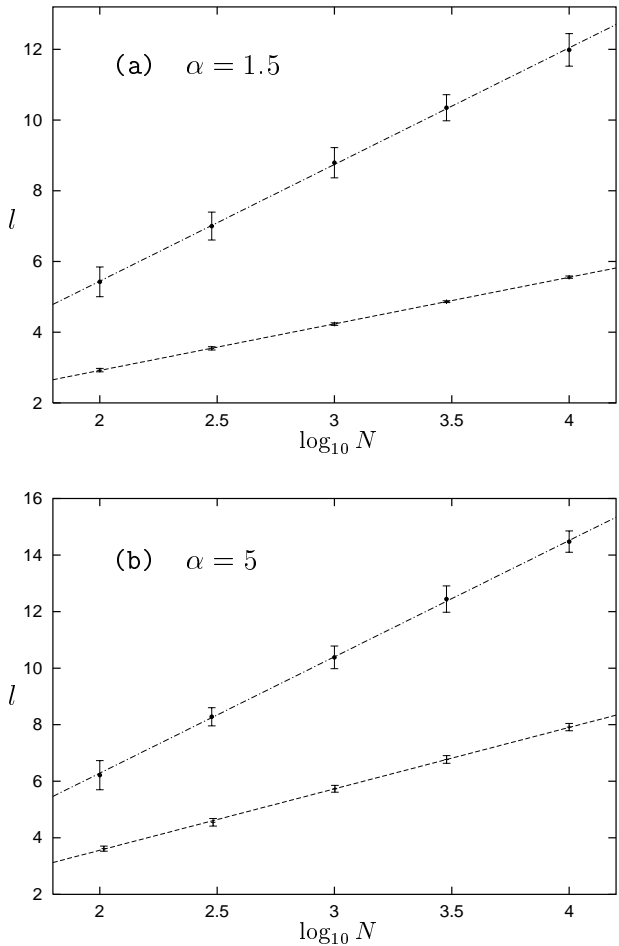,width=7.7in}} 
  \caption{The diameter of a network as a function of the number of sites 
           $N$. Panel (a) corresponds to $\alpha=1.5$ and panel (b) 
           corresponds to $\alpha=5$. The upper lines in each panel are 
           those for $m=1$, the lower lines correspond to $m=3$. Note the 
           logarithmic scale.}
  \label{fig:5}
\end{figure}
\bigskip

In Fig. 5 we plot the mean number of connections between each two nodes of
the network for two different values of $\alpha $ ($\alpha =1.5$ and 
$\alpha=5$) and for the two values $m=1$ and $m=3$ as a function of the network 
size $N$. The algorithm here is trivial: starting from a node (labeled 0) we 
pass to all nodes connected to it (nodes of the first generation, labeled 1), 
then to nodes of the second generation (labeled 2), etc; untill all nodes are
labeled. The mean distance between this node (labeled 0) and any other given
node of the network is then the sum of all values of these labels divided by 
$N-1$. This procedure is repeated for each node, and the overall mean value,
the so-called path diameter of the network ($l$), is evaluated. The error bars of
the figure correspond to the average of the mean diameters over 10 realizations 
of the network. Fig.5 shows that the mean diameter of the network grows linearly 
in $\ln N$, i.e. it shows the typical small-world behavior. This behavior is
preserved for all tested values of $\alpha ;$ the largest value tested was $%
\alpha =45,$ which, for $m=1$, corresponds to a practically sure connection
of a newly introduced node to its nearest neighbor. The high-$\alpha $
networks resemble closely the simple small-world constructions \cite{Watts}.

Let us summarize our findings. We considered a growing network, whose growth
algorithm is based, as in the SF construction, on a preferential attachment
of the newly introduced nodes to the highly connected old ones. However,
here the too long connections are disadvantaged by introducing penalties.
Thus, the probability to connect two nodes separated by a distance $d$ is
proportional to $d^{-\alpha }$, where $\alpha $ is a variable parameter. We
found out that for $\alpha <1$ the degree distribution $P(k)$ decays, as in the 
SF model, like $P(k)\sim k^{3}$, whereas for $\alpha >1$ a stretched exponential 
form $P(k)=a\exp (-bk^{\gamma })$ gives an extremely good description of this 
distribution. On the other hand, the small-world property is preserved at all 
checked values of $\alpha $. 

The authors are grateful to Mr. F. Jasch for fruitful discussions and to Mr. 
M. Obermayr for the valuable technical assistence. Financial support by the 
Fonds der Chemischen Industrie is gratefully acknowledged.

\end{multicols}

\end{document}